\documentclass[aps,prd,preprint,eqsecnum,tightenlines,nofootinbib,showpacs]{revtex4-1}
\pdfoutput=1

\usepackage{amsfonts}
\usepackage{amsmath}
\usepackage{graphicx}
\usepackage{tabularx}
\usepackage{array}

\DeclareMathOperator{\Gr}{Gr}
\DeclareMathOperator{\Li}{Li}

\begin{document}

\title{
{\Large Landau Singularities and Symbology:}\\ One- and Two-loop MHV Amplitudes in SYM Theory
}

\author{Tristan Dennen, Marcus Spradlin and Anastasia Volovich}

\affiliation{
Department of Physics, Brown University, Providence RI 02912, USA
}

\vskip .5 cm
\begin{abstract}
We apply the Landau equations, whose solutions parameterize the locus
of possible branch points, to the one- and two-loop Feynman integrals relevant
to MHV amplitudes in planar $\mathcal{N}=4$ super-Yang-Mills theory. We then
identify which of the Landau singularities appear in the symbols of the amplitudes,
and which do not. We observe that all of the symbol entries in the two-loop MHV
amplitudes are already present as Landau singularities of one-loop pentagon integrals.
\end{abstract}

\maketitle

\tableofcontents

\section{Introduction}

A long-standing goal of the S-matrix program is to be able to construct
formulas for the scattering amplitudes of a quantum field theory based on
general principles and physical constraints.
As amplitudes
are expected (at finite order in perturbation theory)
to be holomorphic functions of the kinematic data,
with isolated poles and branch cuts,
a thorough understanding of their analytic structure is clearly
of paramount importance towards this goal.
In 1959 Landau formulated a simple set of equations~\cite{Landau:1959fi}
whose solutions
parameterize the locus, in the space of kinematic data, where
a given Feynman integral can develop branch points.
Subsequent developments are reviewed in the classic
book~\cite{ELOP}, now celebrating its fiftieth year,
and more recently in Chapter 13 of Ref.~\cite{Sterman:1994ce}.

In a large variety of theories
many of the simplest (and hence best understood) scattering amplitudes
can be expressed in terms of a class of generalized polylogarithm
functions known also as Chen iterated integrals~\cite{Chen}.
Much of the information about the analytic structure
of a function $f$ of this type,
in particular the locations of its logarithmic branch points, is conveniently
encapsulated in an object called the symbol of
$f$~\cite{Goncharov:2009,Goncharov:2010jf}.
The entries of a symbol (often called the ``letters'' of its
``symbol alphabet'') are various
algebraic functions of the kinematic data whose zeros represent
branch point singularities (not necessarily on the principal
sheet) of the corresponding function.

As recently pointed out by Maldacena et~al.~in Ref.~\cite{Maldacena:2015iua},
for amplitudes of generalized polylogarithm type (or, as is commonly
the case, of this type together with an algebraic prefactor),
there should evidently be a close connection between symbol
entries and solutions of the Landau equations.
Specifically,
the symbol entries appearing in any amplitude (in fact even in any
individual Feynman integral)
should be such that their zeros specify
values of external momenta where solutions of the Landau equations exist.
Aspects of this connection have played a role
in Refs.~\cite{Abreu:2014cla,Abreu:2015zaa}, where various one- and
two-loop examples were studied.

Although the Landau/symbol connection ought to be quite general,
in this paper we focus on examples drawn from
$\mathcal{N}=4$ super-Yang-Mills (SYM) theory, specifically the one- and
two-loop MHV amplitudes.  Although the latter involve double pentagon
integrals which are superficially more complicated than the
examples studied in Refs.~\cite{Abreu:2014cla,Abreu:2015zaa},
our analysis benefits from the fact that the kinematic domain
for scattering amplitudes in SYM theory is much simpler than in general
field theories.

As we review in Section 2, in general the Landau equations admit
many families of solutions which are naturally stratified into sets
called the leading Landau singularities (LLS), sub-leading Landau
singularities (SLLS), sub-sub-leading (S${}^2$LLS) etc.
We introduce these abbreviations to carefully distinguish the
Landau singularities of \emph{integrals} from the (closely related)
``leading singularities'' of \emph{integrands}
(see for example Ref.~\cite{ArkaniHamed:2009dn})
which are a key concept in modern work
on the mathematical structure of SYM theory.
To amplify this point:  the leading singularities of an $L$-loop
integrand are the points in the $4L$-dimensional space of
loop integration variables where the integrand, a rational function,
has nonzero residues (the term is also commonly used to refer
to the values of these residues, not their locations);
these leading singularities exist for generic values of the external
kinematics.
In contrast,
the Landau singularities of an $n$-point integral or amplitude are the loci
in the $3(n-5)$-dimensional space of external kinematic data where
the Landau equations admit solutions and where branch points are located.
Note that there is no canonical way to distinguish between the
leading, sub-leading, etc.~Landau singularities of an amplitude---it may
depend on the particular representation of the amplitude in terms of
Feynman integrals. We therefore refer to the LLS, SLLS, etc.~collectively
as ``the Landau singularities.''

Section 2 contains a lightning review of the Landau equations, the
projective geometry used for describing the kinematics of
SYM theory, and the Landau singularities for various simple one-loop
integrals.
In Section 3 we tabulate the Landau singularities of the
chiral pentagon integral, which forms the basic
building block for all one-loop MHV amplitudes.
We find an exact correspondence between the SLLS and S$^2$LLS and the symbol
entries of this integral.
The double pentagon integral, in terms of which all two-loop MHV
amplitudes are expressed, is analyzed in Section 4.  Finally we conclude
with a discussion of our results and questions for further research.

\section{Lightning Review}

\subsection{Landau Singularities}
The Landau equations for a given Feynman integral are a set of kinematic constraints that are necessary for the appearance of a pole or branch point in the integrated function (as a function of external kinematics and masses). There are multiple ways to formulate them (see Ref.~\cite{Landau:1959fi} for a thorough review), although in this paper we stick to the formulation from the Feynman parameterized integral, where the integrand is a function of loop momenta and Feynman parameters, $(\ell_r^\mu,\alpha_i)$:
\begin{align}
\label{feynmanintegral}
I &= c\int \prod_{r=1}^{L} d^D \ell_r \int_{\alpha_i\ge0} d^\nu \alpha\, \delta(1-\sum\nolimits_{i=1}^\nu \alpha_i ) \frac{\mathcal{N}(\ell_r^\mu,p_i^\mu,\ldots)}{\mathcal{D}^\nu}
\end{align}
with
\begin{align}
\mathcal{D} &=  \sum_{i=1}^\nu \alpha_i (q_i^2-m_i^2)\,,
\end{align}
and $c$ a prefactor that will not enter into our discussion. Here, $q_i^\mu$ is the momentum flowing along propagator $i$, $p_i^\mu$ are external momenta, and $\mathcal{N}$ is some numerator function of the kinematic data. There are two distinct situations in which $I$ can develop a singularity or branch point:
\begin{enumerate}
\item The surface $\mathcal{D}=0$ pinches the integration contour in all $(\ell_r,\alpha_i)$ simultaneously. The kinematic locations at which this happens are called ``Leading Landau Singularities" (LLS).
\item The surface $\mathcal{D}=0$ hits the boundary of the integration contour, at $\alpha_i=0$ for some subset of the Feynman parameters, and pinches the contour in all other variables. These are called ``Non-leading Landau Singularities," which we stratify into ``Subleading" (SLLS), ``Sub-sub-leading," (S$^2$LLS) and so forth, according to how many of the $\alpha_i$ are vanishing.
\end{enumerate}
Although we do not review the derivation here, these two situations are captured by the following set of simultaneous equations:
\begin{align}
\sum_{i\in\text{loop}} \alpha_i q_i^\mu & =0 \quad \forall \text{ loops} , \label{kirchhoffrule}\\
\alpha_i (q_i^2-m_i^2) &=0 \quad \forall i. \label{onshellrule}
\end{align}

On the principal sheet, the integration in $\ell_r^\mu$ and $\alpha_i$ in~(\ref{feynmanintegral}) is taken over the real axis, with $\alpha_i\ge0$. Branch points on the principal sheet require the solution to~(\ref{kirchhoffrule}) and~(\ref{onshellrule}) to pinch this contour. When discussing symbol entries, however, we are also interested in branch points on higher sheets, which are exposed by analytically continuing~(\ref{feynmanintegral}) to generic contours. Therefore, throughout this paper we will look more generally for solutions to~(\ref{kirchhoffrule}) and~(\ref{onshellrule}) with $\ell_r^\mu,\alpha_i\in\mathbb{C}$.

\subsection{One-loop Bubbles, Triangles, and Boxes}

The Landau equations~(\ref{kirchhoffrule}) and~(\ref{onshellrule}) are easily solved for one-loop bubble, triangle, and box integrals in four dimensions. Equation~(\ref{onshellrule}) puts all of the propagators on-shell, with no constraints on external kinematics, while the solvability of the loop rule~(\ref{kirchhoffrule}) gives a determinantal constraint after contracting with each of the propagator momenta $q_i^\mu$.

\begin{figure}
	\begin{center}
		\setlength{\tabcolsep}{10pt}
		\begin{tabular}{cc}
			\includegraphics[scale=0.425]{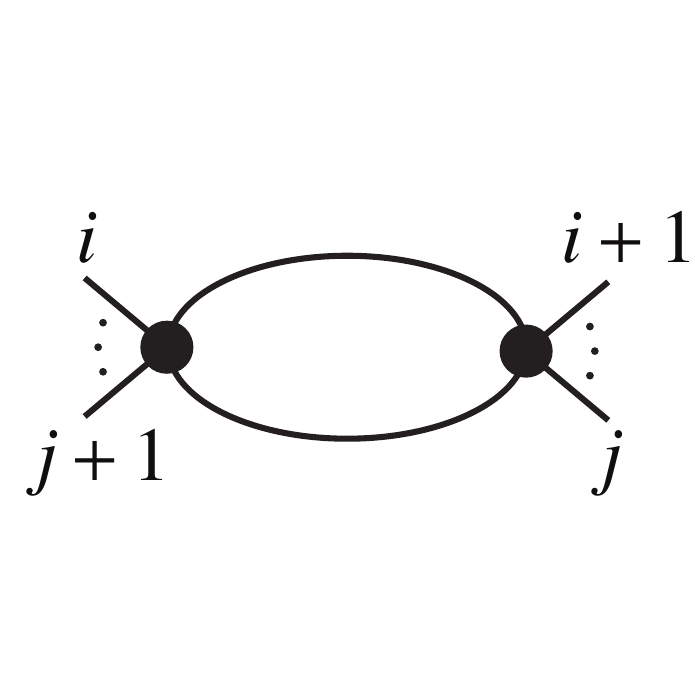}  & \includegraphics[scale=0.425]{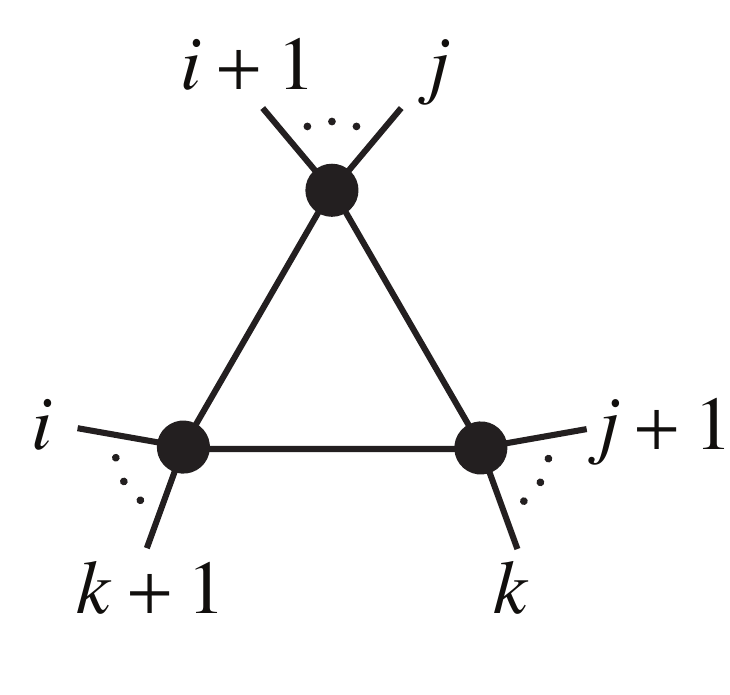} \tabularnewline
			(a) & (b)
		\end{tabular}
	\end{center}
	\caption{The scalar bubble and triangle integrals. All propagators are taken to be massless.}
	\label{fig:bubbletriangle}
\end{figure}
For the bubble and triangle integrals shown in Figure~\ref{fig:bubbletriangle}, the locations of the LLS are given by
\begin{eqnarray}
&\text{Bubble:} \qquad &0 = x_{ij}^2\,,\\
&\text{Triangle:} \qquad &0 = x_{ij}^2 \, x_{jk}^2 \, x_{ik}^2\,, \label{triangleLLS}
\end{eqnarray}
where the region momenta $x_i^\mu$ are defined through $p_i^\mu = x_i^\mu - x_{i{-}1}^\mu$ and $x_{ij}^2 \equiv (x_i-x_j)^2$.
In degenerations of the triangle integral when one or more of the corners is on-shell, (\ref{triangleLLS}) is trivially satisfied for any kinematics. In four dimensions, this is often indicative of the presence of an infrared divergence. More generally, any infrared divergent integral will have either LLS or some S$^k$LLS at unconstrained kinematics.

The scalar box integrals are shown in Figure~\ref{fig:boxintegrals}. As is well known~\cite{Landau:1959fi}, they have the following LLS:
\begin{figure}
	\begin{center}
		\setlength{\tabcolsep}{4pt}
		\begin{tabular}{ccccc}
			\includegraphics[scale=0.425]{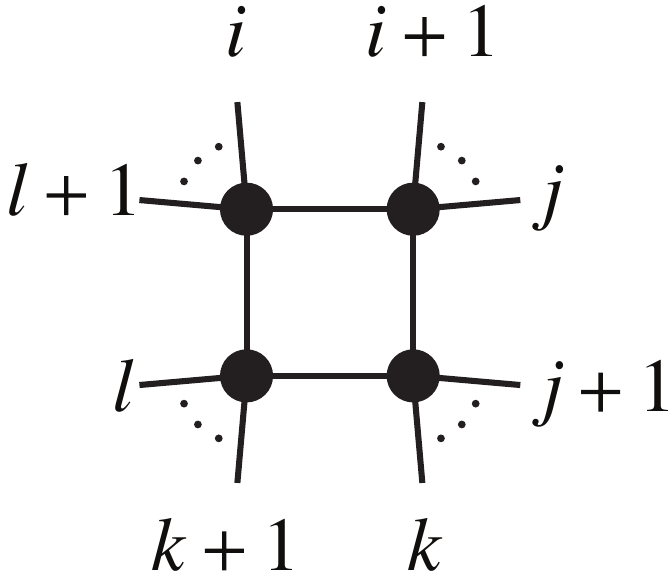} &
			\includegraphics[scale=0.425]{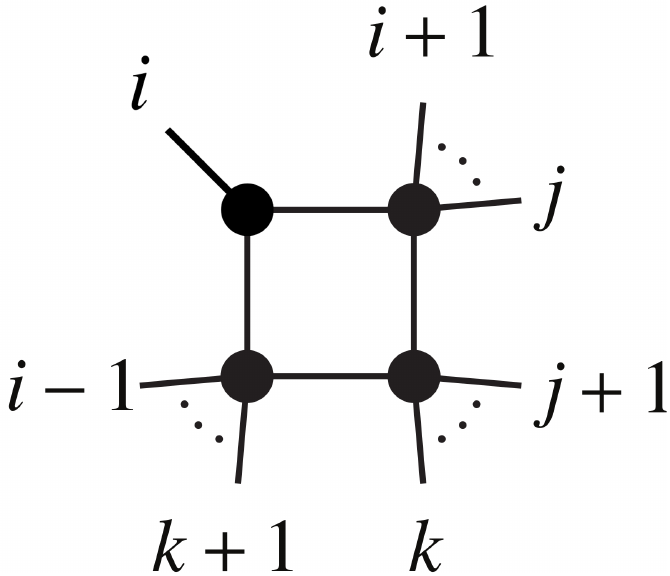} &
			\includegraphics[scale=0.425]{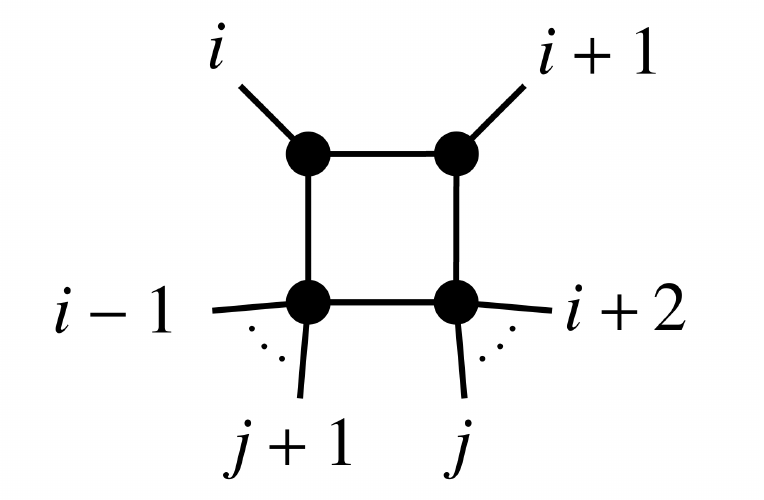} &
			\raisebox{-0.15cm}{\includegraphics[scale=0.425]{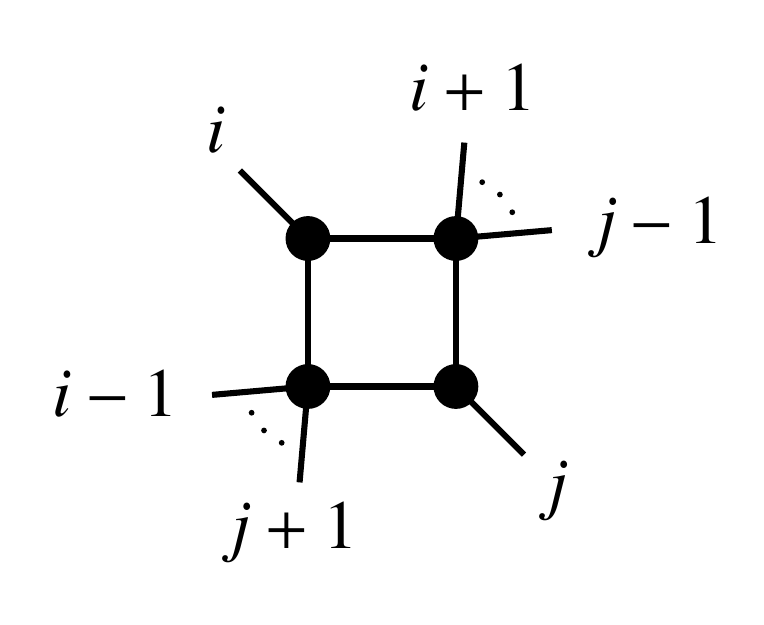}} &
			\raisebox{0.1cm}{\includegraphics[scale=0.425]{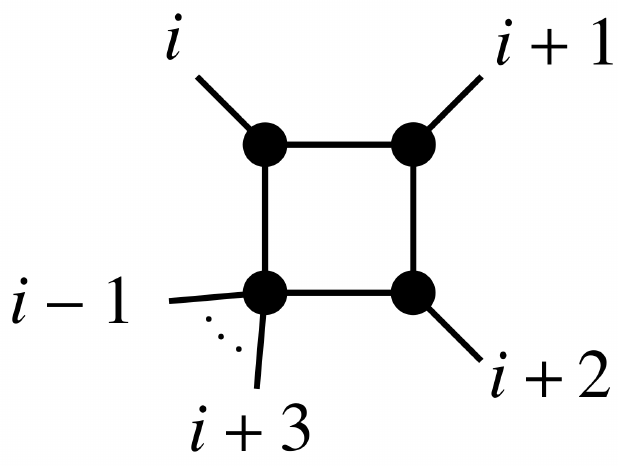}} \tabularnewline
			(a) & (b) & (c) & (d) & (e)
		\end{tabular}
	\end{center}
	\caption{The scalar box integral and its degenerations. All propagators are taken to be massless.}
	\label{fig:boxintegrals}
\end{figure}
\begin{eqnarray}
&\text{Box (a):} \qquad &0 = (x_{ij}^2 x_{kl}^2 - x_{ik}^2 x_{jl}^2 + x_{il}^2 x_{jk}^2)^2 - 4 x_{ij}^2 x_{jk}^2 x_{kl}^2 x_{il}^2 \label{boxALLS} \\
&\text{Box (b):} \qquad &0 = \langle i (i{-}1,i{+}1) (j,j{+}1) (k,k{+}1) \rangle \label{boxBLLS} \\
&\text{Box (c):} \qquad &0 = \langle i{-}1\,i\,i{+}1\,i{+}2 \rangle \langle i\, i{+}1\,j\,j{+}1 \rangle \label{boxCLLS} \\
&\text{Box (d):} \qquad &0 = \langle i\, \overline{j} \rangle \langle \overline{i}\,j \rangle \label{boxDLLS} \\
&\text{Box (e):} \qquad &0 = \langle i{-}1\,i\,i{+}1\,i{+}2 \rangle \langle i\,i{+}1\,i{+}2\,i{+}3 \rangle \label{boxELLS}
\end{eqnarray}
The LLS for (b)--(e) are merely degenerations of (a), and we have written them using momentum twistors (reviewed in the next subsection) for easy comparison with the results of the remainder of this paper. The infrared divergences in boxes (b)--(e) are indicated by trivial vanishing of some triangle-like SLLS. Inspection of explicit formulas for these integrals~\cite{Ellis:2007qk} reveals a number of branch points not captured by the Landau singularities; we will postpone discussion of these other branch points until Section~\ref{sec:Conclusions}.

\subsection{Projective Geometry and Momentum Twistors}

We begin by reviewing some essential facts about
momentum twistor variables~\cite{Hodges:2009hk},
which very conveniently parameterize the kinematic
configuration space for massless scattering
in planar gauge theories, where
(for each color-ordered partial amplitude)
the particles are endowed with a
specific cyclic ordering.
We will see that
momentum twistors enormously simplify the problem of analyzing the
Landau equations in such theories, for the same reason that
they simplify the analysis of leading singularities of the integrand
(see Ref.~\cite{ArkaniHamed:2010gh} for several examples of such
calculations).

Momentum twistors are based on the correspondence between
null rays in (complexified) Minkowski space and points
in twistor space ($\mathbb{P}^3$); or equivalently, between
(complex)
lines in $\mathbb{P}^3$ and
points in Minkowski space.
We use $Z_A, Z_B,$ etc.~to denote points in $\mathbb{P}^3$, which
may be represented using four-component homogeneous
coordinates $Z_A = (Z_A^1, Z_A^2, Z_A^3, Z_A^4)$ subject
to the identification $Z_A \sim t Z_A$ for any nonzero
complex number $t$.
We use $(A,B)$
to denote the line in $\mathbb{P}^3$
containing two given points $Z_A, Z_B$.
Similarly, $(A,B,C)$ denotes the plane containing
the three points $Z_A, Z_B, Z_C$, and
$(A,B,C) \cap (D,E,F)$ denotes the line where this plane
intersects the plane $(D,E,F)$.

Treating the homogeneous coordinates of each momentum twistor
as a vector in $\mathbb{C}^4$, there is a natural
$SL(4,\mathbb{C})$ invariant
denoted by
\begin{equation}
\langle A\,B\,C\,D\rangle \equiv \epsilon_{IJKL} Z_A^I Z_B^J Z_C^K Z_D^L\,.
\end{equation}
We will often be interested to have a geometric understanding of the
locus where
such four-brackets might vanish, which can be pictured in several
ways.
For example, $\langle A\,B\,C\,D\rangle = 0$
only if the two lines $(A,B)$ and $(C,D)$ intersect, or equivalently if
the lines $(A,C)$, $(B,D)$ intersect,
or if the point $A$ lies in the plane $(B,C,D)$, or if the point $C$ lies
on the plane $(A,B,D)$, etc.
Computations of four-brackets involving intersections may be carried
out explicitly with the rule
\begin{equation}
\langle (A,B,C) \cap (D,E,F) \, G\, H\rangle
= \langle A\, B\, C\, G \rangle \langle D\, E\, F\, H\rangle
- \langle A\, B\, C\, H \rangle \langle D\, E\, F\, G\rangle\,.
\end{equation}
This is manifestly antisymmetric under exchanging any
two of the points specifying the plane
$(A,B,C)$, or of any two in $(D,E,F)$, or under exchanging the two planes.
In case the two planes are specified with one common point,
say $F=C$, it is convenient to use the shorthand notation
\begin{equation}
\label{eq:shorthand}
\langle C (A,B) (D,E)(G,H) \rangle \equiv
\langle (A,B,C) \cap (D,E,C)\, G\, H\rangle\,.
\end{equation}
This quantity is also
manifestly
antisymmetric under exchanging $A \leftrightarrow B$,
$D \leftrightarrow E$, or $G \leftrightarrow H$.
Moreover, and less obviously,
is also fully antisymmetric under exchange of any of the three
lines $(A,B)$, $(D,E)$, $(G,H)$.  This may be seen with the help
of the Schouten identity
\begin{equation}
\langle A\,B\,C\,D\rangle Z_E +
\langle B\,C\,D\,E\rangle Z_A +
\langle C\,D\,E\,A\rangle Z_B +
\langle D\,E\,A\,B\rangle Z_C +
\langle E\,A\,B\,C\rangle Z_D = 0\,.
\end{equation}

\begin{figure}
\begin{picture}(117,85)
\put(0,0){\includegraphics[scale=0.6]{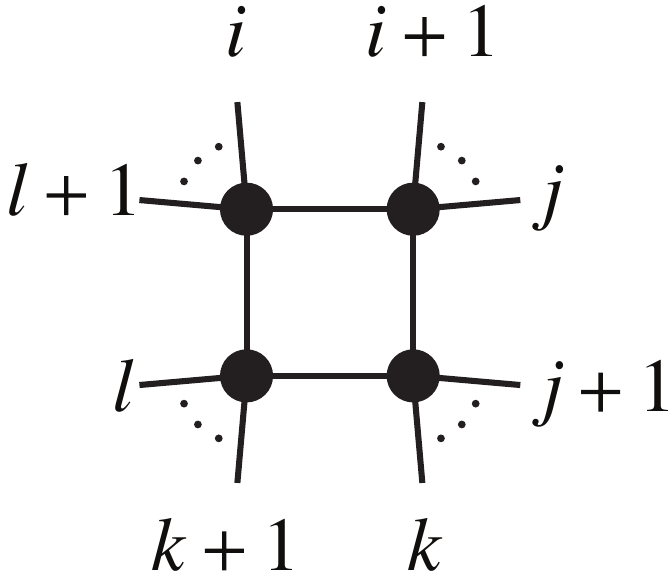}}
\put(25,47){$x_l$}
\put(52,22){$x_k$}
\put(52,72){$x_i$}
\put(77,47){$x_j$}
\end{picture}
\caption{The four-mass box integral.}
\label{fig:fourmass}
\end{figure}

Now we turn to Hodges' construction~\cite{Hodges:2009hk}.
To gain a working understanding of this correspondence
it is instructive to take a look at an explicit example,
such as the one-loop four-mass box integral shown in
Figure~\ref{fig:fourmass}.
In a Feynman diagram it is standard to label each edge according
to the momentum carried by the associated propagator:
$q_1, q_2,$~etc.  In a planar diagram it is equally appropriate
to label the edges by dual momenta (or region momenta):
$x_1, x_2,$~etc.
If two faces labeled $x_a$ and $x_b$ share an edge $q_c$, then the momentum
running along that edge is $q_c = x_a - x_b$ (an overall orientation
must be specified so that the sign of each $q$ in the diagram
is consistent with momentum conservation at each vertex).

The external legs in Figure~\ref{fig:fourmass}
are labeled $1,\ldots,n$ where $n$ is the total
number of particles.
The kinematic data needed to specify the scattering configuration
consists of either the collection $p_1,\ldots,p_n$ of (incoming)
momenta of these particles, or a collection $Z_1,\ldots,Z_n$
of $n$ momentum twistors in $\mathbb{P}^3$.
The former are related to the dual momenta
by $p_a = x_a - x_{a-1}$.  It follows that the four
external face variables labeled
$x_i, x_j, x_k, x_l$ in Figure~\ref{fig:fourmass}
are related to the external momenta by
$(x_i - x_j)^2 = (p_{i{+}1} + p_{i{+}2} + \cdots + p_j)^2$, etc.
Hodges' construction for $n$-point kinematics associates
the point $x_a$ in Minkowski space to the line $(a, a{+}1)$
in $\mathbb{P}^3$ containing the two points $Z_a, Z_{a{+}1}$.
One final standard notation worth mentioning is that
$\bar{a}$ denotes the plane $(a{-}1,a,a{+}1)$.

If $x, y$ are points in Minkowski space associated
to two lines
$(A,B), (C,D)$ in $\mathbb{P}^3$,
then their dual spacetime separation may be computed from the formula
\begin{equation}
(x - y)^2 = \frac{\langle A\,B\,C\,D \rangle}{\langle I \, A\, B \rangle
\langle I\, C\, D\rangle}\,,
\end{equation}
where $I$ is the ``infinity twistor''---the line in twistor space
associated to the dual spacetime point at spatial infinity.
The denominator factors are necessary in order to obtain
the flat Minkowski metric in $x$ space, but we will henceforth
ignore them as they always drop out
of any dual conformal invariant result.

To carry out the integration for the Feynman diagram shown
in Figure~\ref{fig:fourmass} we should associate some
dual momentum $x$ to the interior face of the diagram,
and then integrate the product of the four propagators
$1/(x-x_i)^2$, etc., over $d^4x$.
In momentum twistor space, the point $x$ is represented
by a line $(A,B)$, and we must integrate the product of the
four propagators $1/\langle AB\, i\,i{+}1\rangle$, etc.,
over the space of lines in $\mathbb{P}^3$.
Details about how to construct the measure of integration
may be found in Ref.~\cite{ArkaniHamed:2010gh}.
Note that the singularity in the propagator
$1/(x-x_i)^2$ which arises when $x$ becomes
null separated from $x_i$ is manifested in momentum twistor
space as the singularity in $1/\langle AB\,i\,i{+}1\rangle$
when the line $(A,B)$ intersects the line $(i,i{+}1)$.

\section{One-Loop MHV Amplitudes}
\label{sec:OneLoopMHV}

We now turn our attention to the chiral pentagon
integral, which is the basic building block for one-loop
MHV amplitudes.
This analysis is extremely simple, but it
is instructive to go
through it carefully as a warm-up for the following section.

\subsection{The Chiral Pentagon}

The one-loop MHV amplitude for $n$ particles in SYM theory
may be expressed as~\cite{ArkaniHamed:2010gh}
\begin{equation}
\frac{\mathcal{A}_{\mathrm{MHV}}^{\mathrm{1-loop}}}{\mathcal{A}_{\mathrm{MHV}}^{\mathrm{tree}}}=
\int_{AB}
\sum_{1<i<j<n}
\raisebox{-1.38cm}{\includegraphics[scale=0.425]{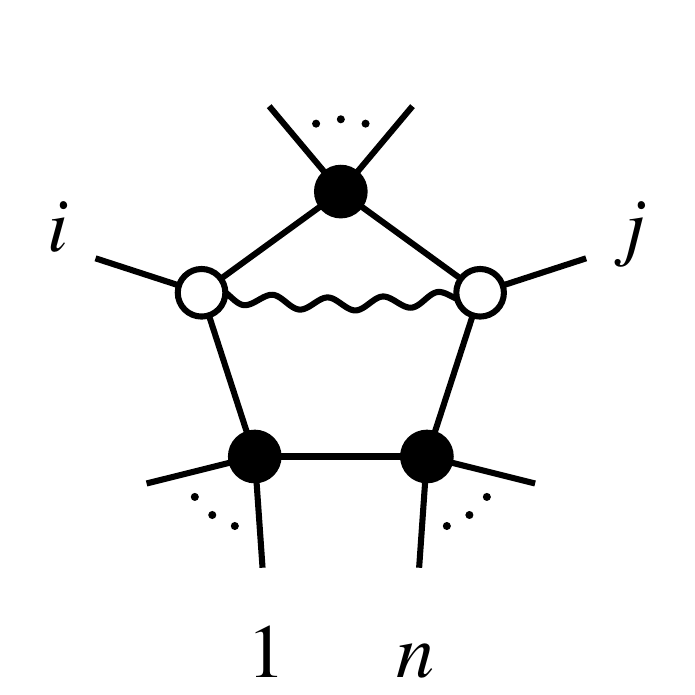}}
\label{eq:oneloopmhv}
\end{equation}
in terms of the chiral pentagon integrand
\begin{equation}
\label{eq:pentagon}
\raisebox{-1.38cm}{\includegraphics[scale=0.425]{mhv_one_loop_single_pentagon.pdf}} =
\frac{\langle AB\,\bar{i} \cap \bar{j} \rangle
\langle i\,j\,n\,1 \rangle}{
\langle AB\,i{-}1\,i\rangle
\langle AB\,i\,i{+}1\rangle
\langle AB\,j{-}1\,j\rangle
\langle AB\,j\,j{+}1\rangle
\langle AB\,n\,1\rangle}\,.
\end{equation}
It is useful to recall several comments
from Ref.~\cite{ArkaniHamed:2010gh}.
First of all, the full integrand
in~(\ref{eq:oneloopmhv}) is cyclic invariant despite the
appearance of the apparently preferred
line $(n,1)$ on the bottom edge of the pentagon;
the formula would be equally valid if $(n, 1)$ were replaced
by $(k, k{+}1)$ for any $k$ (and the summation taken over
$k+1<i<j<k$).

Second, the numerator factors in~(\ref{eq:pentagon}) are specially
chosen so that all of its leading singularities are normalized
to 1.  In fact it would not be inappropriate to say that
half of them are 1 and half of them are 0: the scalar pentagon integral
has twice as many non-zero leading singularities
as the chiral pentagon, but the numerator factors
in~(\ref{eq:pentagon}) vanish on half of them.

Third, for generic $i$ and $j$ the chiral pentagon
integral~(\ref{eq:pentagon}) is infrared finite; the numerator
factor $\langle AB\,\bar{i}\cap \bar{j}\rangle$ softens
the behavior of the integral precisely in the regions of integration
where soft or collinear divergences might appear.
This cancellation fails only for certain boundary terms in the sum
(specifically, when $i=2$ or $j-i=1$ or $j=n-1$) in which case the pentagon
degenerates to an IR divergent box integral.
Henceforth we ignore these degenerate cases since the box
integrals were already reviewed in
the previous section.

An explicit formula for the chiral integral was
obtained in Ref.~\cite{ArkaniHamed:2010gh}:
\begin{equation}
\label{eq:explicitpentagon}
\int_{AB} \raisebox{-1.38cm}{\includegraphics[scale=0.425]{mhv_one_loop_single_pentagon.pdf}} =
\begin{array}{l}
\displaystyle
\Li_2\left(1-u_{n,i-1,i,j}\right)
-\Li_2\left(1-u_{j,n,i,j-1}\right)
+\Li_2\left(1-u_{j,n,i-1,j-1}\right)
\\
\displaystyle
\quad -\Li_2\left(1-u_{i,j-1,n,i-1}\right)
+\Li_2\left(1-u_{i,j-1,j,i-1}\right)
\\
\displaystyle
\quad + \log\left(u_{j,n,i-1,j-1}\right) \log\left(u_{n,i-1,i,j}\right)
\end{array}
\end{equation}
in terms of the dual spacetime cross-ratio
\begin{equation}
u_{i,j,k,l} = \frac{
\langle i\,i{+}1\,j\,j{+}1 \rangle \langle k\,k{+}1\,l\,l{+}1\rangle}
{\langle l\,l{+}1\,j\,j{+}1 \rangle \langle k\,k{+}1 i\,i{+}1\rangle}
= \frac{ x_{ij}^2 x_{kl}^2 }{ x_{lj}^2 x_{ki}^2 }\,.
\end{equation}

From this explicit result we can easily read off the letters appearing
in the symbol of the chiral pentagon.
It is already apparent from~(\ref{eq:explicitpentagon}) that
only the dual spacetime distances $x_{ab}^2$ can appear
in the first entry of the symbol, reflecting
the physically allowed branch points for a scattering amplitude
on the physical sheet~\cite{Gaiotto:2011dt}.
In terms of momentum twistors, the 8 letters that appear
in the first entry are
\begin{equation}
\begin{aligned}
\label{eq:pentagonfirstentry}
&\langle i{-}1\ i\ j{-}1\ j\rangle\,,
\quad
&\langle i{-}1\ i\ j\ j{+}1\rangle\,,
\quad
&\langle i{-}1\ i\ n\ 1\rangle\,,
\quad
&\langle i\ i{+}1\ j{-}1\ j\rangle\,,
 \cr
&\langle i\ i{+}1\ j\ j{+}1\rangle\,,
\quad
&\langle i\ i{+}1\ n\ 1\rangle\,,
\quad
&\langle j{-}1\ j\ n\ 1\rangle\,,
\quad
&\langle j\ j{+}1\ n\ 1\rangle\,.
\end{aligned}
\end{equation}
Six additional letters make an appearance only in the second
entry
of the symbol:
\begin{equation}
\label{eq:pentagonsecondentry}
\begin{aligned}
&\langle \bar{i} j \rangle\,, \quad
&\langle i (i{-}1,i{+}1)(j,j{+}1)(n,1) \rangle\,, \quad
&\langle i (i{-}1,i{+}1)(j{-}1,j)(n,1) \rangle\,, \cr
&\langle \bar{j} i \rangle\,, \quad
&\langle j (j{-}1,j{+}1)(i,i{+}1)(n,1) \rangle\,, \quad
&\langle j (j{-}1,j{+}1)(i{-}1,i)(n,1) \rangle\,.
\end{aligned}
\end{equation}
Four of the letters listed
in~(\ref{eq:pentagonfirstentry})
also can appear in the second entry of the symbol:
$\langle i{-}1\,i\,n\,1\rangle$,
$\langle i\,i{+}1\,n\,1\rangle$,
$\langle j{-}1\,j\,n\,1\rangle$ and
$\langle j\,j{+}1\,n\,1 \rangle$.


\subsection{Landau Singularities}

To find the LLS
of the pentagon, one must solve the Landau equations
\begin{equation}
\begin{aligned}
\label{eq:pentagonlandau}
&
\alpha_1 \langle AB\,i{-}1\,i\rangle
=
\alpha_2 \langle AB\,i\,i{+}1\rangle
=
\alpha_3 \langle AB\,j{-}1\,j\rangle
=
\alpha_4 \langle AB\,j\,j{+}1\rangle
=
\alpha_5 \langle AB\,n\,1\rangle = 0\,, \\
&\alpha_1 (x_{AB} - x_{i{-}1})
+ \alpha_2 (x_{AB} - x_{i})
+ \alpha_3 (x_{AB} - x_{j{-}1})
+ \alpha_4 (x_{AB} - x_{j})
+ \alpha_5 (x_{AB} - x_{n}) = 0 \cr
\end{aligned}
\end{equation}
for all five $\alpha_i$ being nonzero.
The equation on the second line of~(\ref{eq:pentagonlandau})
is content-free in this case---it tells us to find a
vanishing linear combination
of five four-component vectors, which is always possible as long as none of the vectors are zero.

The first four equations tell us to find lines $(A,B)$
which intersect the four given lines $(i{-}1,i)$,
$(i,i{+}1)$, $(j{-}1,j)$ and $(j,j{+}1)$.
For generic $i, j$ (as we have assumed) there are precisely
two solutions to this Schubert problem~\cite{ArkaniHamed:2010gh}:
\begin{equation}
\label{eq:schubert1}
(A,B) = (i,j) \qquad \text{or} \qquad (A,B) = \bar{i} \cap \bar{j}\,.
\end{equation}
Geometrically this is clear:  we can either take $(A,B)$ to be
the line $(i,j)$ which contains the two points $Z_i, Z_j$, or
we can take $(A,B)$ to be the intersection of the
planes $(i{-}1,i,i{+}1)$ and $(j{-}1,j,j{+}1)$.

It only remains to solve the equation $\langle AB\,n\,1\rangle = 0$,
but upon plugging in the solution~(\ref{eq:schubert1}) this
becomes a constraint on the external kinematics:
\begin{flalign}
\label{eq:pentagonLLS}
\text{\bf (LLS)} &&
\langle i\,j\,n\,1 \rangle \langle n\,1\,\bar{i} \cap \bar{j} \rangle = 0\,.&&
\end{flalign}
To conclude:  solutions of the Landau equations~(\ref{eq:pentagonlandau})
with all
$\alpha_i \ne 0$ exist only on the locus in kinematic space
where~(\ref{eq:pentagonLLS}) is satisfied.

The SLLS of the pentagon are found by solving the Landau
equations~(\ref{eq:pentagonlandau})
with four of the five $\alpha$'s being nonzero.  Each case
amounts to a degeneration of the pentagon to a box, so we can
simply
transcribe
the results of the previous solution.  For vanishing
$\alpha_1$, $\alpha_2$, $\alpha_3$, $\alpha_4$ or $\alpha_5$
we find respectively that the SLLS lie on the loci:
\begin{flalign}
\text{\bf (SLLS)}
&&
\begin{array}{l}
\langle j (j{-}1,j{+}1)(i,i{+}1)(n,1) \rangle= 0\,, \\
\langle j (j{-}1,j{+}1)(i{-}1,i)(n,1) \rangle= 0\,, \\
\langle i (i{-}1,i{+}1)(j,j{+}1)(n,1) \rangle= 0\,, \\
\langle i (i{-}1,i{+}1)(j{-}1,j)(n,1) \rangle= 0\,, \\
\langle \bar{i} j \rangle \langle i  \bar{j} \rangle= 0\,.
\end{array}
&&
\label{eq:pentagonSLLS}
\end{flalign}

The S${}^2$LLS are given by solutions of~(\ref{eq:pentagonlandau}) with
only three nonzero $\alpha$'s, which correspond to degenerations of the
pentagon to various triangles.
The four non-trivial cases, arising from three-mass triangles,
are
\begin{flalign}
\text{\bf (S${}^2$LLS)}
&&
\begin{array}{l}
\langle i{-}1\ i\ j{-}1\ j \rangle \langle j{-}1\ j\ n\ 1
\rangle \langle n\ 1\ i{-}1\ i \rangle= 0\,,\\
\langle i\ i{+}1\ j{-}1\ j \rangle \langle j{-}1\ j\ n\ 1
\rangle \langle n\ 1\ i\ i{+}1 \rangle= 0\,,\\
\langle i{-}1\ i\ j\ j{+}1 \rangle \langle j\ j{+}1\ n\ 1
\rangle \langle n\ 1\ i{-}1\ i \rangle= 0\,,\\
\langle i\ i{+}1\ j\ j{+}1 \rangle \langle j\ j{+}1\ n\ 1
\rangle \langle n\ 1\ i\ i{+}1 \rangle= 0\,.
\end{array}
&&
\label{eq:pentagonSSLLS}
\end{flalign}
Degenerations which lead to two-mass triangles give solutions
of the Landau equations for all kinematics, as reviewed in the previous
section.  These singularities, in the case of the scalar pentagon,
are indicative of the soft and collinear IR singularities of the integral.
We know however that (for generic $i, j$, as always) the numerator
factor in~(\ref{eq:pentagon}) eliminates these singularities.

We could go one step further, down to bubbles, but this provides
no new information:  all bubbles are either fully singular or have
Landau singularities on the vanishing loci of brackets
which already appear
in~(\ref{eq:pentagonSSLLS}).

\subsection{Summary}

We have tabulated all Landau singularities of the pentagon integral.
Some sufficiently degenerate singularities exist for all
kinematics.
Often such singularities are indicative of IR divergences, but
we know that for this particular integral (and for generic $i, j$)
these are canceled
by the numerator factor in~(\ref{eq:pentagon}).
Let us emphasize that except for appealing to this fact,
the analysis of the previous section applies to the scalar
pentagon integral just as well as the chiral integral, since
the Landau equations by definition only know about the
propagator structure of a diagram.

The singularities that exist only on various nontrivial submanifolds
of kinematic space are indicated in
equations~(\ref{eq:pentagonLLS}), (\ref{eq:pentagonSLLS})
and~(\ref{eq:pentagonSSLLS}).
Upon comparison with equations~(\ref{eq:pentagonfirstentry})
and~(\ref{eq:pentagonsecondentry}) we notice
a striking pattern:  sub-sub-leading Landau
singularities~(\ref{eq:pentagonSSLLS}) exist
only on the loci where the leftmost symbol
entries~(\ref{eq:pentagonfirstentry}) vanish, while
sub-leading singularities~(\ref{eq:pentagonSLLS}) appear
on a different set of loci corresponding to the locations
where the second-entry symbol entries~(\ref{eq:pentagonsecondentry})
vanish.
(However let us not forget that~(\ref{eq:pentagonsecondentry})
only lists the {\it new} letters which start to appear in
the second entry.)

What about the LLS, which lives
on the locus
$\langle i\,j\,n\,1 \rangle \langle n\,1\,\bar{i} \cap \bar{j} \rangle = 0$?
This quantity indeed makes an appearance
in the overall prefactor in the scalar pentagon integral,
which evaluates (see for example Ref.~\cite{Bern:1993kr}) schematically to
$1/{\Delta}$ times a transcendental function of uniform weight 2,
where $\Delta \propto \langle i\,\bar{j} \rangle \langle \bar{i}\,j\rangle
\langle i\,j\,n\,1 \rangle \langle n\,1\,\bar{i} \cap \bar{j}\rangle$.
The chiral pentagon, however, is a {\it pure} integral: as is
evident from~(\ref{eq:explicitpentagon}), it evaluates to a
transcendental function with no algebraic prefactor.
This cancellation is achieved by the carefully chosen numerator
of the chiral pentagon, and it is intimately connected with the
fact that its integrand has ``unit leading singularities,''
as emphasized in Ref.~\cite{ArkaniHamed:2010gh}.

Evidently this is a happy example where there is a very clear
separation between the LLS, which tell us only about the overall
algebraic singularities of the amplitude (and which actually end
up washed out by the fact that the leading singularities of the
integrand are normalized to 1), and the sub-(sub-)leading
Landau singularities which probe past the prefactor
and into the symbol.
Let us recall that MHV amplitudes
are expected to evaluate to  pure transcendental functions,
with no algebraic prefactors (other than the
tree-level MHV amplitude
indicated on the left-hand side of~(\ref{eq:oneloopmhv}))
to all orders in
perturbation theory~\cite{ArkaniHamed:2012nw}.

\section{Two-Loop MHV Amplitudes}
\label{sec:TwoLoopMHV}

We now turn our attention to the chiral double pentagon
integral, which is the basic building block for two-loop
MHV amplitudes.

\subsection{The Chiral Double Pentagon}

The two-loop MHV amplitude for $n$ particles in SYM
theory may be expressed as~\cite{ArkaniHamed:2010gh}
\begin{equation}
\label{eq:twoloopmhv}
\frac{\mathcal{A}_{\mathrm{MHV}}^{\mathrm{2-loop}}}{\mathcal{A}_{\mathrm{MHV}}^{\mathrm{tree}}}=
\int_{AB} \int_{CD}
\underset{\substack{i<j<k<l<i}}{\frac{1}{2}\text{{\Huge$\sum$}}\phantom{\frac{1}{2}\!\!}}\hspace{-0.2cm}\raisebox{-1.38cm}{\includegraphics[scale=0.425]{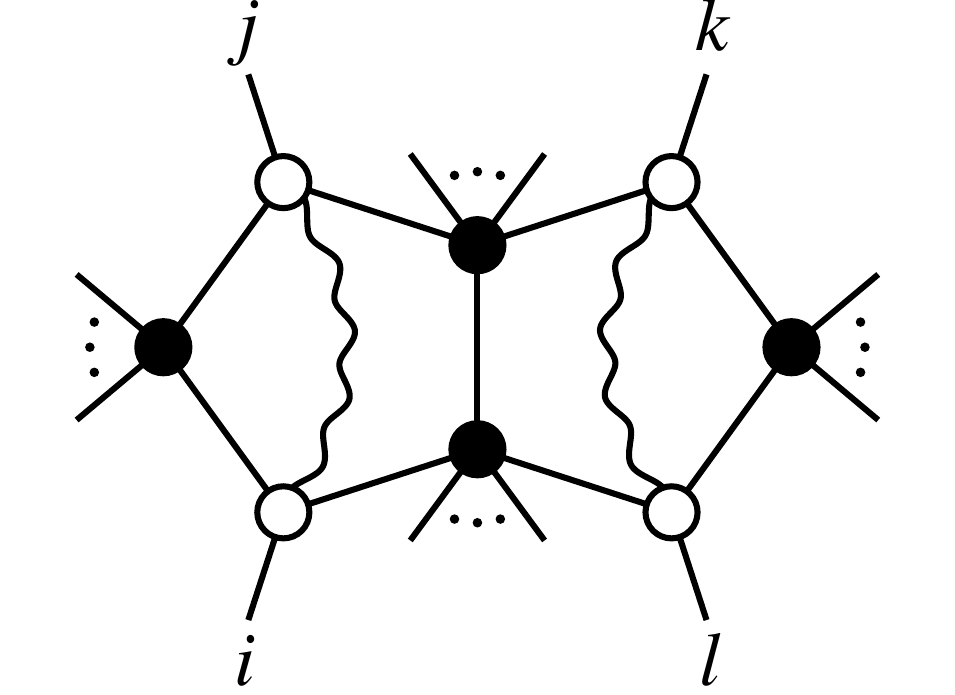}}
\end{equation}
in terms of the chiral double pentagon integrand
\begin{equation}
\label{eq:doublepentagon}
\raisebox{-1.38cm}{\includegraphics[scale=0.425]{two_loop_mhv_integrand.pdf}}=
\begin{array}{l}
\displaystyle
\frac{\langle i\,j\,k\,l\rangle}{\langle ABCD\rangle}
\\
\displaystyle
\quad \times
\displaystyle
\frac{\langle AB\, \bar{i} \cap \bar{j}\rangle}
{\langle AB\, i{-}1\,i\rangle
\langle AB\,i\,i{+}1\rangle
\langle AB\,j{-}1\,j\rangle
\langle AB\,j\,j{+}1\rangle}
\\
\displaystyle
\quad \times
\frac{
\langle CD\, \bar{k} \cap \bar{l} \rangle}{
\langle CD \,k{-}1\,k \rangle
\langle CD \,k\,k{+}1\rangle
\langle CD \,l{-}1\,l\rangle
\langle CD\,l\,l{+}1\rangle}\,.
\end{array}
\end{equation}

The numerator factors in~(\ref{eq:doublepentagon}) serve the same
purpose as in the one-loop pentagon discussed
in the previous section.  Each of the two
nontrivial numerator factors vanishes on half of the leading singularities
of the scalar double pentagon integrand; their product is non-zero
on one quarter of them.  The integrand is normalized to have
residue 1 on these leading singularities.
The numerator factors also suppress the soft/collinear divergences,
rendering the integral finite for
generic $i,j,k,l$.

Explicit analytic results for the chiral double pentagon integral
have been obtained only for the special case
$l = k + 2 = j + 3 = i + 5$
at $n=6$~\cite{Drummond:2010mb}.
However it is expected that for generic $i,j,k,l$ the integral
is expressible as a generalized polylogarithm with a symbol
alphabet similar to that described
in~(\ref{eq:pentagonfirstentry}) and~(\ref{eq:pentagonsecondentry}).
Specifically, the letters appearing in the first entry of the
symbol are expected on general physical grounds~\cite{Gaiotto:2011dt}
to be
\begin{equation}
\label{eq:letter1}
\langle a\,a{+}1\,b\,b{+}1\rangle, \qquad a, b \in \{i{-}1,i,j{-}1,j,
k{-}1,k,l{-}1,l\}\,.
\end{equation}
In the second entry we expect to start seeing additional letters
of the type
\begin{equation}
\label{eq:letter2}
\langle a \bar{b} \rangle\,, \qquad
\langle a(a{-}1,a{+}1)(c,c{+}1)(d,d{+}1)\rangle
\end{equation}
for
$a,b \in \{i,j,k,l\}$ and $c,d \in \{i{-}1,i,j{-}1,j,
k{-}1,k,l{-}1,l\}$.

We are less certain about the symbol alphabet for the third and fourth
entries of the chiral double pentagon integral.
For guidance we rely on the symbol of the full two-loop $n$-point
MHV amplitude~(\ref{eq:twoloopmhv}),
which was determined in Ref.~\cite{CaronHuot:2011ky} and which contains,
in its third and fourth entries, letters of the form
\begin{equation}
\label{eq:letter3}
\langle a\, a{+}1\, b\, c\rangle
\quad
\text{and}
\quad
\langle a\, a{+}1\, \bar{b} \cap \bar{c} \rangle\,.
\end{equation}
We cannot rule out the possibility that individual chiral double pentagon
integrals might have an even larger symbol alphabet, with nontrivial
cancellation in the sum~(\ref{eq:twoloopmhv}).
Indeed some cancellation is known to occur: the final entry of the symbol
of an MHV amplitude
is always of the form $\langle a\,\bar{b}\rangle$~\cite{CaronHuot:2011ky},
but we do not expect this to be true for each individual chiral
double pentagon
integral.

\subsection{Landau Singularities}

To find the leading Landau singularities of the double pentagon
we must solve for $AB$, $CD$ which put all nine propagators on-shell.
The loop rule~(\ref{kirchhoffrule}) plays no
role in this case for the same reason discussed
under~(\ref{eq:pentagonlandau}).
The eight propagators
\begin{equation}
\label{eq:eightprops}
\begin{aligned}
\langle AB\,i{-}1\,i\rangle =
\langle AB\,i\,i{+}1\rangle =
\langle AB\,j{-}1\,j\rangle =
\langle AB\,j\,j{+}1 \rangle &= 0\cr
\langle CD\,k{-}1\,k\rangle =
\langle CD\,k\,k{+}1\rangle =
\langle CD\,l{-}1\,l\rangle =
\langle CD\,l\,l{+}1 \rangle & =0
\end{aligned}
\end{equation}
are put on shell just as in~(\ref{eq:schubert1}),
by
\begin{equation}
(A,B) = (i,j)~\text{or}~\bar{i} \cap \bar{j} \quad
\text{and}
\quad
(C,D) = (k,l)~\text{or}~\bar{k} \cap \bar{l}\,.
\label{eq:schubert2}
\end{equation}
It remains only to set the ninth propagator
$\langle ABCD \rangle$
to zero, but in light of~(\ref{eq:schubert2})
this becomes a constraint on the external kinematics:
\begin{flalign}
\label{eq:doubleLLS}
\text{\bf (LLS)} &&
\langle i\ j\ k\ l \rangle
\langle i\ j\ \bar{k} \cap \bar{l} \rangle
\langle \bar{i} \cap \bar{j}\ k\ l\rangle
\langle \bar{i} \cap \bar{j}\ \bar{k} \cap \bar{l} \rangle = 0\,.
&&
\end{flalign}
The Landau equations for the double pentagon admit a leading
solution only on the locus in kinematic space
where~(\ref{eq:doubleLLS}) is satisfied.

At sub-leading order there are two possible topologies.  If
the $\alpha$ associated to the internal propagator
$\langle ABCD\rangle$ is set to zero then we simply get
two kissing boxes of type (d) in
Figure~\ref{fig:boxintegrals}, whose Landau singularities, according
to~(\ref{boxDLLS}), lie on the intersection of
\begin{flalign}
\text{\bf (SLLS)} &&
\langle i \bar{j} \rangle
\langle \bar{i} j \rangle = 0\quad
\text{and} \quad
\langle k \bar{l} \rangle \langle \bar{k} l \rangle = 0\,.
&&
\end{flalign}

On the other hand if the $\alpha$ associated to one of the
eight propagators displayed in~(\ref{eq:eightprops}) is set
to zero then the diagram degenerates to a pentagon-box.
Suppose for example that we collapse the edge $(i,i{+}1)$.
Then the $AB$ loop becomes a 3-mass box whose leading Landau
singularity is at
\begin{equation}
\langle j(j{-}1,j{+}1)(i,i{-}1) CD \rangle = 0
\end{equation}
according to~(\ref{boxBLLS}).
Meanwhile $CD$ is determined by the same four equations
on the second line of~(\ref{eq:eightprops}) to be
$(C,D) = (k,l)$ or $(C,D) = \bar{k} \cap \bar{l}$.
This particular sub-leading Landau singularity therefore
lives on the locus
\begin{equation}
\langle j(j{-}1,j{+}1)(i,i{-}1) (k,l)\rangle
\langle j(j{-}1,j{+}1)(i,i{-}1)\, \bar{k} \cap \bar{l}\rangle = 0\,.
\end{equation}
Altogether there are a total of eight such sub-leading
singularities:
\begin{flalign}
\text{\bf (SLLS)}
&&
\begin{array}{l}
\langle j (j{-}1,j{+}1) (i{-}1,i) (k,l) \rangle
 \langle j (j{-}1,j{+}1) (i{-}1,i) \ \bar{k} \cap \bar{l}\rangle
= 0\,,\\
 \langle j (j{-}1,j{+}1) (i,i{+}1) (k,l) \rangle
 \langle j (j{-}1,j{+}1) (i{-}1,i) \ \bar{k} \cap \bar{l}\rangle
= 0\,,\\
 \langle i (i{-}1,i{+}1) (j{-}1,j) (k,l) \rangle
 \langle j (j{-}1,j{+}1) (i{-}1,i) \ \bar{k} \cap \bar{l}\rangle
= 0\,,\\
 \langle i (i{-}1,i{+}1) (j,j{+}1) (k,l) \rangle
 \langle j (j{-}1,j{+}1) (i{-}1,i) \ \bar{k} \cap \bar{l}\rangle
= 0\,.
\end{array}
&&
\label{eq:doublepentagonSLLS}
\end{flalign}
and four more of the same type but with $ij$
and $kl$ exchanged.

At sub-sub-leading order we can have a triangle-pentagon or
a double-box.  The former can be obtained by further collapsing
the $AB$ integral from a box, as we have just discussed, down
to a triangle.  Using~(\ref{triangleLLS})
we find
a total of eight non-trivial triangle-pentagon singularities:
\begin{flalign}
\text{\bf (S$^2$LLS)}
&&
\begin{array}{l}
\langle i\ i{+}1\ j{-}1\ j \rangle
\langle j{-}1\ j\ k\ l \rangle
\langle k\ l\ i\ i{+}1\rangle
\langle j{-}1\ j\ \bar{k} \cap \bar{l} \rangle
\langle \bar{k} \cap \bar{l}\ i\ i{+}1\rangle
= 0\,,\\
\langle i{-}1\ i\ j{-}1\ j \rangle
\langle j{-}1\ j\ k\ l \rangle
\langle k\ l\ i{-}1\ i\rangle
\langle j{-}1\ j\ \bar{k} \cap \bar{l} \rangle
\langle \bar{k} \cap \bar{l}\ i{-}1\ i\rangle
= 0\,,\\
\langle i\ i{+}1\ j\ j{+}1 \rangle
\langle j\ j{+}1\ k\ l \rangle
\langle k\ l\ i\ i{+}1\rangle
\langle j{-}1\ j\ \bar{k} \cap \bar{l} \rangle
\langle \bar{k} \cap \bar{l}\ i{-}1\ i\rangle
= 0\,,\\
\langle i{-}1\ i\ j\ j{+}1 \rangle
\langle j\ j{+}1\ k\ l \rangle
\langle k\ l\ i{-}1\ i\rangle
\langle j{-}1\ j\ \bar{k} \cap \bar{l} \rangle
\langle \bar{k} \cap \bar{l}\ i{-}1\ i\rangle
= 0\,,
\end{array}
&&
\end{flalign}
and again four more of the same type but with
$ij$ and $kl$ exchanged.
We have included on this list only those degenerations which
lead to a three-mass triangle diagram.  As discussed earlier,
two-mass triangles admit solutions to the Landau equations for
generic kinematics.

Next we consider the sub-sub-leading singularities of double
box type.  Suppose
we collapse the edges $(i,i{+}1)$ and $(l{-},l)$.
The three-mass box on the right has a leading Landau singularity on the
locus $\langle k(k{-}1,k{+}1)(l,l{+}1)\,AB\rangle = 0$.
Taking this condition together with the three remaining propagators
\begin{equation}
\langle AB\,i{-}1\,i\rangle = \langle AB\,j{-}1\,j\rangle
= \langle AB\,j\,j{+}1\rangle = \langle k(k{-}1,k{+}1)(l,l{+}1)AB\rangle = 0\,,
\end{equation}
we see that the left box is also of three-mass type, specified
by the four lines $(i{-}1,i)$, $(j{-}1,j)$, $(j,j{+}1)$
and, with the help of~(\ref{eq:shorthand}),
$ k(k{-}1,k{+}1)(l,l{+}1) = \bar{k} \cap (k,l,l{+}1)$.
This three-mass box has its
leading Landau singularity on the locus
\begin{equation}
\langle j(j{-}1,j{+}1)(i{-}i,i)\ \bar{k} \cap (k,l,l{+}1) \rangle
= \langle \bar{j} \cap (i{-}1,i,j)\ \bar{k} \cap (k,l,l{+}1)\rangle = 0\,.
\end{equation}
The double pentagon has a total of 16 sub-subleading singularities of
this type, given by:
\begin{flalign}
\label{eq:complicated}
\text{\bf (S$^2$LLS)}
&&
\begin{array}{l}
\langle \bar{i} \cap (i,j{-}1,j) \ \bar{l} \cap (k,k{+}1,l) \rangle = 0\,, \\
\langle \bar{i} \cap (i,j,j{+}1) \ \bar{l} \cap (k,k{+}1,l) \rangle = 0\,, \\
\langle \bar{i} \cap (i,j{-}1,j) \ \bar{l} \cap (k{-}1,k,l) \rangle = 0\,, \\
\langle \bar{i} \cap (i,j,j{+}1) \ \bar{l} \cap (k{-}1,k,l) \rangle = 0\,,
\end{array}
&&
\end{flalign}
and the same with $i \leftrightarrow j$ or $k \leftrightarrow l$.

At sub$^3$-leading order we can have singularities
coming from triangle-box topologies.  There are a total of 32
degenerations of this type involving a three-mass triangle
and a three-mass box, with Landau singularities at the following
loci in kinematic space:
\begin{flalign}
\text{\bf (S$^3$LLS)}
&&
\begin{array}{l}
\langle i\ i{+}1\ j{-}1\ j\rangle
\langle k (k{-}1,k{+}1)(l,l{+}1)(j{-}1,j)\rangle
\langle k (k{-}1,k{+}1)(l,l{+}1)(i,i{+}1)\rangle = 0\,, \\
\langle i\ i{+}1\ j\ j{+}1\rangle
\langle k (k{-}1,k{+}1)(l,l{+}1)(j,j{+}1)\rangle
\langle k (k{-}1,k{+}1)(l,l{+}1)(i,i{+}1)\rangle = 0\,, \\
\langle i{-}1\ i\ j{-}1\ j\rangle
\langle k (k{-}1,k{+}1)(l,l{+}1)(j{-}1,j)\rangle
\langle k (k{-}1,k{+}1)(l,l{+}1)(i{-}1,i)\rangle = 0\,, \\
\langle i{-}1\ i\ j\ j{+}1\rangle
\langle k (k{-}1,k{+}1)(l,l{+}1)(j,j{+}1)\rangle
\langle k (k{-}1,k{+}1)(l,l{+}1)(i{-}1,i)\rangle = 0\,,
\end{array}
&&
\end{flalign}
then four more like these but with $(l{-}1,l)$ instead of
$(l,l{+}1)$, and then another eight given by
taking $k \leftrightarrow l$, and finally another sixteen
given by those described so far but with $ij$ and $kl$ exchanged.

We could go further down to degenerations involving
bubbles, or the triangle-triangle graph, but we will stop here
as we have already encountered in the Landau
singularities seen so far all of the four-brackets
which are known to appear in the symbol of the two-loop MHV amplitudes.
In fact we expect that all triangle-triangle and bubble-$*$ diagrams
to be either fully singular or to have Landau singularities on
the vanishing loci of four-brackets which already appear.

\subsection{Summary and Discussion}

The chiral double pentagon integral has quite a long laundry
list of Landau singularities.  Actually, as we confess in the following
section, there are additional classes of singularities that we have
not even looked at.  It is however clear that all of the
symbol entries of the
two-loop MHV amplitudes,
shown in~(\ref{eq:letter1}), (\ref{eq:letter2}) and~(\ref{eq:letter3}),
do vanish on loci where the Landau equations admit solutions.
It is an important question for future work to discover whether
it is possible to make a stronger statement explaining why
the various other Landau singularities do not seem to manifest
themselves as symbol entries.

In particular
the S${}^2$LLS of double box type, shown
in~(\ref{eq:complicated}), involve more complicated
four-brackets than those which appear in actual two-loop MHV amplitudes.
These brackets bear a resemblance to some of the more complicated
cluster $\mathcal{A}$-coordinates which appear in the $\Gr(4,n)$
Grassmannian cluster algebra that is apparently
relevant for scattering amplitudes
in planar SYM theory~\cite{Golden:2013xva} (see in particular
eq.~(6.18)).
We comment on the intriguing possibility
of a deeper connection between Landau singularities and cluster algebras
at the end of the next section.

\section{Conclusion and Caveats}
\label{sec:Conclusions}

Motivated by the observation of Ref.~\cite{Maldacena:2015iua} that there
should be a close connection between symbol
entries and solutions of the Landau equations, in
this paper we initiated a study of the Landau singularities of
Feynman integrals relevant to one- and two-loop MHV amplitudes in
planar SYM theory.  On general physical grounds it is expected that
a quantity may appear in the symbol of some Feynman integral only if
the locus where it vanishes corresponds to some Landau singularity; i.e.~only
if the Landau equations admit a solution on that locus.

At one loop we found a surprising crisp statement: the sub-sub-leading
Landau singularities of the pentagon
live on the loci where the first entries of the
symbols of the MHV
amplitudes vanish, while the sub-leading Landau singularities live on
the loci where the second entries vanish.
The leading Landau singularity appears as an overall pole in the
scalar pentagon integral but is canceled by the numerator factor
in the chiral pentagon integral~(\ref{eq:pentagon}) whose role
is to unit normalize the leading singularity of the integrand.

The analysis of the two-loop Landau equations is more complicated.
Although we did verify
that all of the symbol entries~(\ref{eq:letter1}), (\ref{eq:letter2})
and~(\ref{eq:letter3}) which are known to appear in two-loop MHV
amplitudes vanish on loci where the Landau equations admit solutions,
we also found numerous additional solutions which seem not to have
a direct connection with the symbol alphabet.
We do not have a satisfactory explanation that accounts for
the presence of some Landau singularities as symbol letters, while
others are excluded.

However, it is very intriguing to note,
by comparing~(\ref{eq:letter1})--(\ref{eq:letter3}) to~(\ref{eq:pentagonLLS})--(\ref{eq:pentagonSSLLS}), that all of the symbol
entries of the \emph{two-loop} MHV amplitudes
were already present
in our classification of the Landau singularities of the
\emph{one-loop} chiral pentagon integral!
The two-loop chiral double pentagon certainly has a large list of
new Landau singularities, but none of these make any appearance
in the symbol of the two-loop amplitude.
It is tempting to speculate that this is connected with the
fact that for MHV amplitudes, all possible Yangian invariants
are exhausted already at one loop~\cite{ArkaniHamed:2012nw};
this means that the $L$-loop MHV integrand has no new leading
singularities beyond those which are present at one loop.
Clearly it would be interesting to extend our analysis
to the NMHV amplitudes and beyond.  (The two-loop NMHV
integrands were written in Ref.~\cite{ArkaniHamed:2010gh}, and
all two-loop integrands were constructed in Ref.~\cite{Bourjaily:2015jna}.)

It is clear that we have only taken the first steps in exploring
a potentially fruitful connection between symbology and the Landau
equations.
In the next few paragraphs we discuss several important caveats
regarding our analysis.
Properly understanding each of these caveats is an interesting, new
research direction.

(1) Perhaps the most important caveat is that the Landau equations themselves
are completely insensitive to the numerator structure of a Feynman
integral; they only identify the locus in kinematic space where
a singularity {\emph{may}} occur.  Yet we know very well that numerator
factors in the integrand of SYM theory are not just some afterthought
thrown on top of a product of propagators, they are part and parcel
of the deep mathematical structure of the integrand.
At one loop we were able, with a little bit of handwaving, to explain
away the absence of certain Landau singularities in the symbol by
appealing to the numerator factor.  Clearly it would be important
to have a more systematic understanding of the role of
numerator factors in SYM theory vis \`a vis the Landau singularity analysis.

(2) In the analyses of Sections~\ref{sec:OneLoopMHV} and~\ref{sec:TwoLoopMHV}, we have ignored an entire class of branch points called ``second-type singularities"~\cite{Fairlie:1962aa,Fairlie:1962ab,ELOP}. These are not solutions to the Landau equations~(\ref{kirchhoffrule}) and~(\ref{onshellrule}). Instead, they are pinch singularities of~(\ref{feynmanintegral}) at infinite loop momenta, for which~(\ref{kirchhoffrule}) and~(\ref{onshellrule}) need to be modified.
However, due to dual conformal symmetry,
there is no invariant notion of ``infinite momentum'' in planar SYM theory,
so these second-type singularities should be absent except when using an infrared regulator that breaks the dual conformal symmetry.
We therefore expect these singularities to be
intimately tied to the infrared divergences of SYM theory, and we believe they are completely captured by the BDS subtraction prescription~\cite{Bern:2005iz} (which can be implemented in a dual conformal preserving way~\cite{Alday:2009zm}).
In any case, recall that the chiral double pentagon integral
is actually finite (and hence dual conformal
invariant) except for certain boundary cases in $i, j, k, l$.

In non-dual-conformally invariant theories, second-type singularities are ubiquitous and not directly tied to IR divergences. As a simple example, consider the three-mass triangle integral which evaluates explicitly to~\cite{'tHooft:1978xw, Lu:1992ny, Bern:1993kr}
\begin{align}
I_3 &= \frac{i}{\sqrt{\Delta}} \sum_{i=1}^{3} \Big( \Li_2(x_i) - \Li_2(1/x_i)\Big), \qquad
x_i = \frac{p_1^2+p_2^2+p_3^2-2p_i^2+ \sqrt{-\Delta}}{p_1^2+p_2^2+p_3^2-2p_i^2 -\sqrt{-\Delta}}\,,
\end{align}
where
\begin{align}
\Delta &= -(p_1^2)^2-(p_2^2)^2-(p_3^2)^2 + 2\, p_1^2 \, p_2^2 + 2\, p_2^2\,p_3^2 +2\, p_1^2 \, p_3^2  \,
\end{align}
is the triangle Gram determinant. This integral is IR finite yet has an algebraic branch point at a second-type singularity, when $\Delta=0$.

(3) We were fortunate in our analysis of the one- and two-loop pentagons
that the Landau singularities we found always live on the vanishing locus
of simple geometric quantities expressible in terms of four-brackets of
twistors, and that the symbol entries of the corresponding amplitudes
were similarly expressible simply in terms of four-brackets.
In general, we should not expect the Landau equation analysis to tell
us what the symbol entries \emph{are}, only where they vanish.
The three-mass triangle presented above illustrates a more complicated
case:  the LLS of the triangle are located at $p_i^2 = 0$, and
there is a second-type singularity at $\Delta = 0$, but the arguments
of the dilogarithms, and hence the symbol entries, are much more
complicated quantities.
Closer inspection of these arguments reveals that they do vanish (or become infinite), as required, precisely on the leading Landau singularity locus
where~(\ref{triangleLLS}) is satisfied.
However one would have been hard-pressed to guess the exact form of
these symbol entries just knowing that the Landau singularities are
at $p_i^2 = 0$ and $\Delta = 0$.  In many cases at one loop, the actual symbol entries are understood geometrically~\cite{Goncharov:1996aa, Spradlin:2011wp},
but there is a shortage of appropriate data on higher-loop integrals.
We can hope that planar
SYM theory is sufficiently special that it is possible
to make a more direct connection
between Landau singularities and symbol alphabets.

(4) We should emphasize again that our two-loop Landau equation
analysis focused on an individual chiral double pentagon integral,
which we compared to the symbol alphabet of the full amplitude
given by the sum~(\ref{eq:twoloopmhv}).  Therefore we must
acknowledge the possibility of nontrivial cancellation in the sum,
such that some symbol entries appearing in individual double pentagon
integrals might drop out of the full amplitude.

(5) Finally, although we consider this very implausible, we should
acknowledge the possibility that the amplitude, or individual integrals,
might have singularities which are only present in beyond-the-symbol
terms.

One of the more speculative
motivations for this work was the tantalizing possibility that there
might exist a physical principle which could dictate the symbol
alphabet for a given integral or amplitude;
often in actual calculations one must start with an ansatz.
We are still very far from understanding whether such a principle
exists or may be useful, but it is worthwhile to explore this
question since the payoff could be so great.
In particular, all evidence available to date is consistent
with the hypothesis that all $n$-point MHV and NMHV amplitudes in planar
SYM theory have symbol alphabets consisting of the set of
cluster coordinates on the $\Gr(4,n)$ Grassmannian cluster
algebra~\cite{Golden:2013xva}. This has been exploited in some of the recent work calculating
remainder and ratio functions~\cite{Dixon:2011pw,Dixon:2011nj,Dixon:2013eka,Dixon:2014voa,Dixon:2014xca,Dixon:2014iba,Drummond:2014ffa,Dixon:2015iva}.
It would be very interesting to explore the interplay
between the physical principle that the singularities of amplitudes
are encoded in the Landau equations, and the
empirically observed mathematical
fact that their singularities are dictated by cluster algebras.

\section*{Acknowledgments}

We are grateful to J.~Maldacena for discussions.
This work was supported by the US Department of Energy
under contract DE-SC0010010 Task A (MS) and
Early Career Award DE-FG02-11ER41742 (AV), as well as
by the Simons Investigator
Award of AV.

\end{document}